\documentclass[twocolumn,resetfootnote,longbib]{aastex7}

\newcommand\multimoon{\texttt{MultiMoon}}
\newcommand\jt{$J_2$}
\newcommand\ct{$C_{22}$}

\usepackage[utf8]{inputenc}
\usepackage{tipa}
\usepackage{combelow}
\usepackage{savesym}
\savesymbol{tablenum}
\usepackage{siunitx}
\restoresymbol{SIX}{tablenum}
\usepackage{breqn}
\usepackage{tabularx}
\usepackage{amsmath}
\usepackage{paralist}
\usepackage{graphicx}
\usepackage{subcaption}
\usepackage{placeins}

\shorttitle{Beyond Point Masses. VI. Typhon--Echidna}
\shortauthors{Proudfoot et al.}

\begin{document}

\title{Beyond Point Masses. VI. Spin-Orbit Evolution of the Centaur Binary Typhon--Echidna}

\author[orcid=0000-0002-1788-870X,sname='Proudfoot']{Benjamin Proudfoot}
\affiliation{Florida Space Institute, University of Central Florida, 12354 Research Parkway, Orlando, FL 32826, USA}
\email[show]{benp175@gmail.com}

\author[orcid=0000-0002-8296-6540, sname='Grundy']{Will Grundy} 
\affiliation{Northern Arizona University, Department of Astronomy \& Planetary Science, PO Box 6010, Flagstaff, AZ 86011, USA}
\email{}

\author[orcid=0000-0003-1080-9770, sname='Ragozzine']{Darin Ragozzine} 
\affiliation{Brigham Young University Department of Physics \& Astronomy, N283 ESC, Brigham Young University, Provo, UT 84602, USA}
\email{darin_ragozzine@byu.edu}

\begin{abstract}
Only three binaries have been identified among the Centaur population. Because their perihelia are significantly closer than those of other trans-Neptunian binaries (TNBs), these systems allow a detailed look at tight binaries in the broader TNO population and provide critical insight into the disruption of binaries as they enter the Centaur population. Using recent and archival \textit{Hubble Space Telescope} (HST) observations, along with Keck data, we present a spin-orbit study of Typhon--Echidna. We find that the binary’s mutual orbit is inconsistent with a Keplerian orbit; more detailed non-Keplerian fits show that the mutual orbit is rapidly precessing. We measure Typhon's dynamical oblateness, $J_2$, at $\sim10\sigma$ confidence and find that Typhon's rotation pole is $\gtrsim20\degr$ misaligned with the binary's mutual orbit. Assuming Typhon has a triaxial shape, our results, combined with rotational light curves and thermal measurements from the literature, suggest ellipsoidal semi-axes of $a=93^{+8}_{-6}$ km, $b=84^{+6}_{-6}$ km, and $c=65^{+9}_{-8}$ km. We further investigate the observational consequences of the complex spin-orbit dynamics, including light curve alteration by axial precession of Typhon and substantial changes to the system's mutual event season. Based on the system’s dynamically excited state, we suggest a recent encounter with a giant planet may have substantially altered the system, potentially consistent with a binary in an early stage of disruption. This hypothesis can be tested with resolved photometric observations of the system. Our investigation highlights how non-Keplerian dynamics enhances our understanding of TNB systems and motivates ongoing observations of TNBs with astrometry, photometry, and stellar occultations.

\end{abstract}


\section{Introduction}
\label{sec:intro}

Since the first discovery of a small trans-Neptunian binary (TNBs) system in 2002 \citep{2002Natur.416..711V}, TNB studies have consistently enabled deep characterization of this distant population of small bodies \citep[e.g.,][]{grundy2019mutual,noll2020trans}. TNBs are ubiquitous among the trans-Neptunian population, making up large fractions of all trans-Neptunian object (TNO) subpopulations \citep[e.g.,][]{stephens2006detection,noll2007binaries}. With TNOs being the source population of the Centaurs \citep{1997Sci...276.1670D,di2025dynamics}, the short-lived small bodies on unstable orbits interacting with the giant planets\footnote{Here, we use the term Centaur to refer to any non-resonant object with perihelion within Neptune's orbit \citep{elliot2005deep,khain2020dynamical}.}, one would naturally expect that Centaurs are also host to numerous binaries \citep{sickafoose2025near}. Instead, only three Centaur binaries are known today, Typhon--Echidna \citep{noll2006discovery}, Ceto--Phorcys \citep{2007Icar..191..286G}, and 2012 KU$_{50}$ \citep{2021CBET.5075....1C}, although the last is only marginally within Neptune's aphelion at its own perihelion. Recent large surveys with the \textit{Hubble Space Telescope} (HST) have confirmed the lack of Centaur binaries \citep{LiHSTprogram}, although binaries with closer separations and/or faint secondaries could have evaded detection.

Although the lifetimes of Centaurs have been studied extensively \citep{gladman1990fates}, comparatively little is known about the lifetimes of binary Centaurs \citep[for a review, see][]{sickafoose2025near}. Explaining the dichotomy between the TNO and Centaur binary fractions has largely focused on the disruption of binaries during their transition from scattered disk/TNO orbits to Centaur orbits \citep{noll2006discovery,araujo2018journey}. As TNOs transition to become Centaurs, bodies begin to sustain close encounters with the giant planets, leading to disruption \citep{brunini2014dynamical}. Based on the two known binary Centaurs, studies have estimated that binary lifetimes in the Centaur population are of order $\sim$Myrs \citep{araujo2018journey}. With a population of three---or two with published orbit solutions---it is difficult to say if they are representative of the original source population. Indeed, a number of observational biases limit the comparability of the TNB and Centaur populations \citep{sickafoose2025near}.

The lack of Centaur binaries is also supported by the apparent lack of binaries among the Jupiter family comets (JFCs). About one-third of Centaurs become JFCs during their inward evolution \citep{levison1997kuiper}. To date, no JFC binaries have been detected, although bilobate shapes are common \citep{keller2015insolation}.

Given the unknown nature of the TNO--Centaur transition for binaries, detailed investigation into the known TNO binaries is critical. The two with published orbit solutions, Typhon and Ceto, are in some ways similar, but in others quite dissimilar. Although both are among the tightest known TNBs, Ceto--Phorcys appears to be tidally evolved with a mutual orbit eccentricity $e<0.01$ \citep{2007Icar..191..286G}, while Typhon--Echidna has $e=0.53$ \citep[][this work]{grundy200842355}. Typhon's excited orbital state is suggestive of recent encounters with giant planets, reflecting an epoch of transient perturbations on the way to binary disruption. The eccentricity of Typhon's mutual orbit, along with its tight orbit, also make it a prime candidate for further detailed physical characterization with non-Keplerian orbit fitting, which can reveal unique information about Typhon's shape and rotation state \citep{proudfoot2024bpm2}.

Non-Keplerian orbit fitting relies on incorporating the effects of an object's non-spherical shape into models of a body's satellite \citep[for an introduction, see][]{proudfoot2024bpm2}. With enough astrometric data on a body's satellite, non-Keplerian orbit fitting can constrain the distribution of angular momentum within a binary system, allowing a determination of mass ratios, rotation poles, and orbital/axial precession. Non-Keplerian effects are beginning to become detectable in many TNB systems \citep{proudfoot2024bpm2}, and the constraints they reveal have already begun to inform us about the shapes, spins, and even interior structure of TNOs \citep{proudfoot2024bpm3,nelsen2025beyond,proudfoot2025beyond}. 

In this study, we focus on physical characterization of the Typhon--Echidna binary with non-Keplerian orbit fitting. Based on two decades of data from HST and Keck (see Section \ref{sec:obs}), we fit a spin-orbit model to astrometric data of the binary which provides a substantial improvement in fit quality when compared to a Keplerian orbit model (see Section \ref{sec:methods}). We then discuss the implications of our spin-orbit model, showing that both orbital and axial precession are substantial (Section \ref{sec:sporb}), inferring the triaxial shape of Typhon (Section \ref{sec:shape}), and placing constraints on the Typhon--Echidna mass ratio (Section \ref{sec:mr}). With such substantial spin-orbit precession, we explore how key observables, like light curves (Section \ref{sec:lc}) and the occurrence of mutual events (Section \ref{sec:me}), are modified under the influence of precession. We finally explore the tidal evolution of the system (Section \ref{sec:tide}) and hypothesize that the system provides a rare glimpse into the ongoing disruption of a binary (Section \ref{sec:origin}), before concluding (Section \ref{sec:conclusions}). 

\section{Observations}
\label{sec:obs}

\begin{deluxetable*}{cccCCCC}
\tablewidth{\textwidth}
\tablecaption{Observed Astrometric Positions of Echidna\label{tab:observations}}
\tablehead{
Julian Date & Date & Telescope/Instrument & \Delta \alpha \cos{\delta} & \sigma_{\alpha} & \Delta \delta & \sigma_{\delta} \\
 & & & ('') & ('') & ('') & ('')
}
\startdata
2453755.91818 & 2006/01/20 & HST/ACS-HRC & -0.08727 & 0.00290  & -0.08172 & 0.00290  \\
2453781.05226 & 2006/02/14 & HST/ACS-HRC & +0.06642  & 0.00190  & -0.15552 & 0.00190  \\
2453785.57234 & 2006/02/19 & HST/ACS-HRC & +0.13570   & 0.00190  & -0.08515 & 0.00190  \\
2454044.56851 & 2006/11/05 & HST/ACS-HRC & -0.00355 & 0.00190  & -0.13523 & 0.00190  \\
2454091.95205 & 2006/12/22 & HST/ACS-HRC & +0.10178  & 0.00190  & +0.00323 & 0.00190  \\
2456373.97397 & 2013/03/22 & Keck/NIRC2  & -0.08799 & 0.00300   & -0.02190 & 0.00300   \\
2458223.70160 & 2018/04/15 & HST/WFC3    & +0.00171  & 0.03169 & +0.00155  & 0.02739 \\
2460780.08354 & 2025/04/14 & HST/WFC3    & -0.06733 & 0.00217 & -0.01540  & 0.00290  \\
2460798.43523 & 2025/05/02 & HST/WFC3    & -0.04332 & 0.00369 & -0.02273 & 0.00207 \\
\enddata
\tablecomments{All astrometry is referenced to the position of Typhon. Observations from 2006 were described in \citet{grundy200842355}.}
\end{deluxetable*}

The orbit fitting we present in this paper is based on approximately two decades of precise relative astrometry from both ground- and space-based observatories. Some of these observations have previously been published \citep{grundy200842355}, but in this work, we present four new observations of the Typhon--Echidna system from Keck and HST. Keck observations at one epoch in 2013 were acquired using the Keck laser guide star adaptive optics system \citep{wizinowich2006} with the NIRC2 narrow camera\footnote{\url{https://www2.keck.hawaii.edu/inst/nirc2}}. These observations followed the same observing and data reduction set up as those conducted in previous surveys of TNBs with Keck \citep[e.g.,][and references therein]{grundy2019mutual}. 

Three epochs of HST imaging have also been acquired of the Typhon system, with a single epoch in 2018 (HST program 15344, PI: Jewitt, see \citealp{LiHSTprogram}) and two epochs in 2025 (HST program 17707, PI: Proudfoot). All images were acquired in the wide F350LP filter, to provide the deepest observations possible within a single HST orbit. Calibrated observations downloaded from MAST were analyzed using a PSF fitting routine, which uses TinyTim model PSFs \citep[][]{krist201120}. These routines have significant heritage and have been used extensively in the TNB orbit fitting literature \citep[e.g.,][etc.]{grundy200842355,grundy2011five,grundy2019mutual}.

All astrometric observations are shown in Table \ref{tab:observations}. All HST observations, both new and archival, are publicly available at DOI:\dataset[10.17909/3ece-tn50]{http://dx.doi.org/10.17909/3ece-tn50}.

\section{Orbit Fitting}
\label{sec:methods}

\subsection{Orbit Model}
Our orbit fitting was completed with the \multimoon{} orbit fitting package \citep{ragozzine2024beyond}. \multimoon{} poses the orbit fitting problem as a Bayesian parameter inference exercise and uses a Markov Chain Monte Carlo (MCMC) sampling algorithm to explore the posterior distribution for each parameter. In addition, \multimoon{} can perform orbit fits with either a Keplerian orbit model or an $N$-quadrupole orbit integrator, which accounts for both $N$-body effects as well as quadrupole-order shape effects from nonspherical components. The integrator also includes the effects of back torques---torques from a secondary on the primary's nonspherical shape---which allows for exploration of complex spin-orbit effects like axial precession. For more details on \multimoon{} and how it is used, see \citet{ragozzine2024beyond} and \citet{proudfoot2024bpm2}.

Based on Keplerian orbit fitting of the astrometry in Table \ref{tab:observations}, we found that the dataset was inconsistent with a Keplerian orbit at the $\sim12\sigma$ level, indicating a high likelihood that the system would display non-Keplerian effects. Indeed, Typhon--Echidna is one of the tightest known TNBs, which maximizes the precession rate of the system. Likewise, its eccentric mutual orbit makes any precession readily observable. \added{Typhon is also a good candidate for having a substantially nonspherical shape, based on its size and apparent lack of tidal evolution \citep{porter2012kctf}.}

To fully model the system in a non-Keplerin spin-orbit framework, we ran a series of \multimoon{} orbit fits using \multimoon's $N$-quadrupole integrator. In these fits, we allow Typhon's gravitational harmonic---$J_2R^2$---to vary. Unlike much of the asteroid binary literature, we explicitly fit $J_2R^2$ instead of $J_2$ alone, since TNO sizes are quite often poorly measured and subject to a variety of assumptions. In our fits, we neglect the effects of the sectoral harmonic $C_{22}R^2$. \ct{} has little effect outside of spin-orbit resonances \citep[][]{proudfoot2021prolate}, and altering it does not change the angular momentum of a body \citep{1995geph.conf....1Y}. 

Since we want to model the spin-orbit effects in the system, a precise spin period and overall size for Typhon are needed to determine the amount of angular momentum in Typhon's spin. \citet{thirouin2010short} suggested a 9.67 hour single-peaked rotation period for the system, with a low-amplitude. If Typhon is triaxial, the double peaked rotation solution is favored with a period of 19.34 hours. Unfortunately, the confidence in the 9.67 hour (or 19.34 hour double-peaked) solution was low, with many other solutions at periods $<1$ day. With only a tentative measurement, we allow the rotation period to vary as a free parameter within our modeling, although we limit it to between 5-20 hours, a range of likely rotation periods \citep[see][]{thirouin2014rotational}. In our resulting fits, goodness-of-fit is $\sim$uncorrelated with rotation period. For rotation periods $>20$ hours, we find good orbit solutions, but the required mass ratio between Typhon and Echidna for valid solutions is implausibly small (see Section \ref{sec:mr} for further details).

To set Typhon's overall size, \multimoon{} requires input of Typhon's $c$-axis. For this, we look to thermal observations of the system; \citet{santos2012tnos} used \textit{Spitzer} and \textit{Herschel} observations to determine the combined system effective diameter and, assuming equal albedos and a 1.3 mag difference between Typhon and Echidna, derived an equivalent diameter of Typhon of $D=162\pm7$ km. For simplicity, we input $c=81$ km. We note that the rotational angular momentum of Typhon, which is important for determining the spin-orbit evolution of the system, is independent of $c$ \citep{ragozzine2024beyond}, and thus our selection of Typhon's $c$-axis has little effect on the dynamics of the system. In some preliminary orbit fits, we allowed $c$ to vary as a free parameter and found that it had no effect on fit quality or retrieved parameters. 

We note that in fits with larger ranges of rotation periods and floating $c$, all parameters except the Typhon--Echidna mass ratio were $\sim$uncorrelated with either the rotation period or $c$.

Inclusion of Echidna's shape and rotation provide only very small perturbations to the system as a whole. The angular momentum of Echidna is $L_{\rm spin}\approx I_{E} \omega_{E} = \frac{2}{5}M_{E}R_{E}^2\omega_{E} = \frac{8}{15}\pi R_{E}^5\rho_{E}\omega_{E}$. With a size ratio between Typhon and Echidna of $\sim0.5$ and $L_{\rm spin}\propto R^5$, the angular momentum contained within Echidna is only a few percent of that in its orbital motion or in Typhon's rotation. Hence, we neglect Echidna's rotation properties in our model, which provides substantial reduction in the number of parameters that need to be explored. 

Other than the above specifications, \multimoon{} fits were run similarly to past TNB orbit fits in terms of priors, initial walker positions, walker burn-in, under-performing walker pruning, etc. In total, our final MCMC run ran with 960 walkers for 25,000 burn-in steps and 20,000 sampling steps. For further details on how our fits are conducted, see \citet{proudfoot2024bpm2} and \citet{nelsen2025beyond}.

\begin{figure*}
    \centering\includegraphics[width=0.99\textwidth]{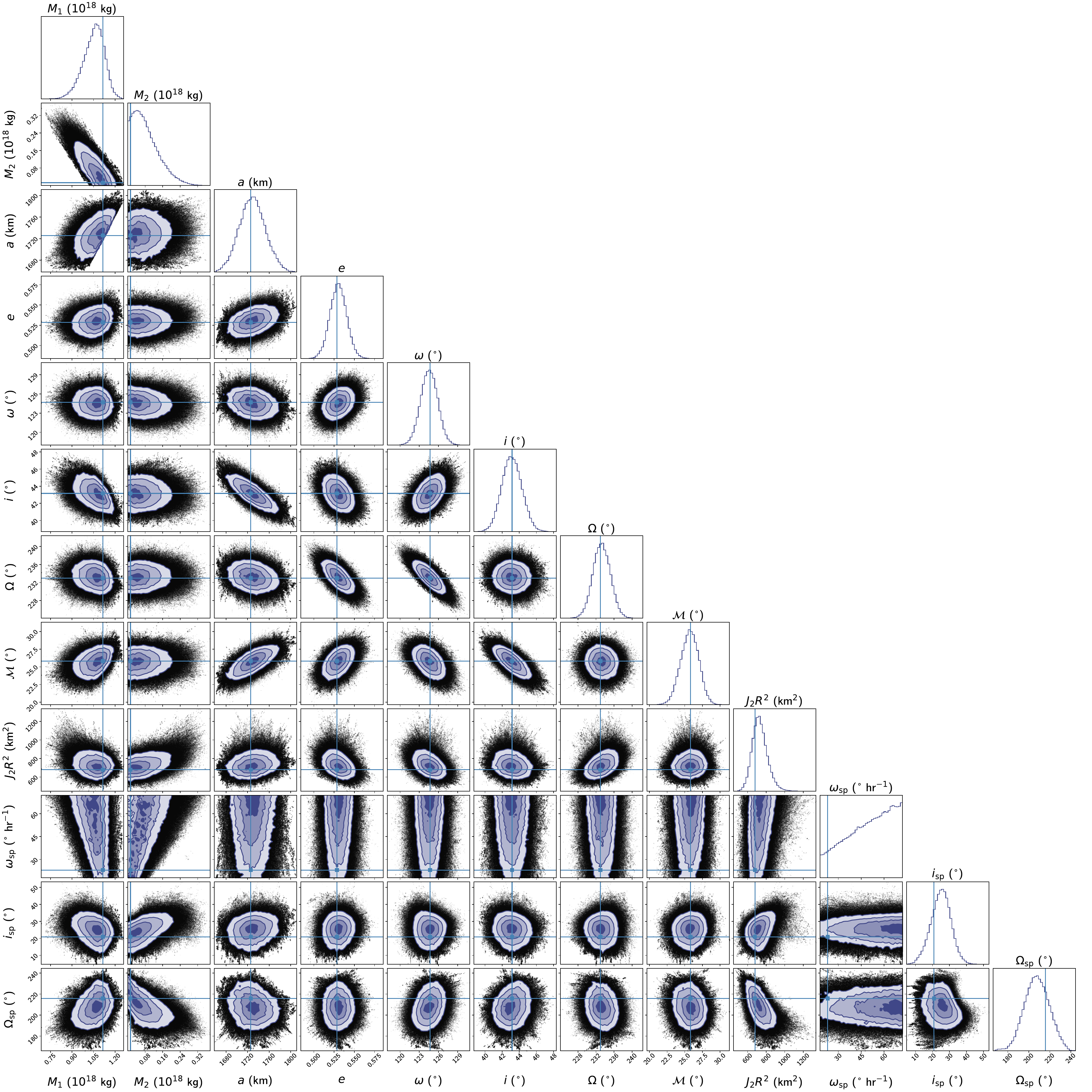}
    \caption{A corner plot showing the posterior distribution from our non-Keplerian orbit fit. At the top of each column, marginal (one-dimensional) posterior distributions for each parameter are shown. Beneath, two-dimensional joint posterior distributions are shown for every pair of parameters. Contours on the joint distributions show the 0.5, 1, 1.5, and 2 $\sigma$ confidence intervals. Horizontal and vertical lines show the location of the best-fit parameter set.}
    \label{fig:corner}
\end{figure*}

\begin{deluxetable}{lcl}
\tablecaption{Non-Keplerian Orbit Solution for Typhon--Echidna\label{tab:orbitfit}}
\tablewidth{0pt}
\tablehead{
\colhead{Parameter} & \colhead{} & \colhead{Posterior}
}
\startdata
Typhon mass ($10^{18}$ kg)                          & $M_{t}$            & $1.06^{+0.06}_{-0.08}$        \\
Echidna mass ($10^{18}$ kg)                        & $M_{e}$            & $<0.14$        \\
Semi-major axis (km)                                  & $a$              & $1729^{+24}_{-24}$            \\
Eccentricity                                          & $e$              & $0.53^{+0.01}_{-0.01}$        \\
Inclination (deg)                              & $i$              & $43.1^{+1.2}_{-1.1}$          \\
Argument of periapse (deg)                     & $\omega$         & $124.6^{+1.3}_{-1.3}$         \\
Nodal longitude (deg)          & $\Omega$         & $233.2^{+1.9}_{-1.9}$         \\
Mean anomaly (deg)                    & $\mathcal{M}$    & $25.7^{+1.3}_{-1.3}$          \\
Typhon $J_2$ harmonic (km$^2$) & $J_2R^2$         & $726^{+87}_{-73}$             \\
Typhon rotation rate (deg hr$^{-1}$)           & $\omega_{\rm sp}$ & $>33$              \\
Axis obliquity (deg)                  & $i_{\rm sp}$         & $25^{+6}_{-6}$                \\
Axis precession (deg)                 & $\Omega_{\rm sp}$    & $208^{+12}_{-12}$             \\
\hline
System mass ($10^{18}$ kg)                            & $M_{\rm sys}$        & $1.13^{+0.05}_{-0.05}$        \\
Echidna-Typhon mass ratio                             & $q$              & $0.068^{+0.078}_{-0.046}$     \\
Typhon obliquity (deg)    & $\epsilon$       & $23.8^{+5.1}_{-4.6}$          \\
Orbit period (days)                         & $P_{\rm orb}$      & $18.990^{+0.004}_{-0.003}$ \\
\added{Typhon $J_2$} & \added{$J_2$} & \added{$0.114^{+0.025}_{-0.021}$} \\
Apsidal precession rate (deg/yr)               & $\dot{\varpi}$   & $3.3^{+0.6}_{-0.6}$           \\
Nodal precession rate (deg/yr)                 & $\dot{\Omega}$   & $-4.4^{+0.4}_{-0.4}$          \\
Orbit pole R.A. (deg) & $\alpha_{\rm orb}$ & $171.2^{+2.3}_{-2.3}$ \\
Orbit pole decl. (deg) & $\delta_{\rm orb}$ & $56.3^{+1.2}_{-1.2}$ \\
Typhon rotation pole R.A. (deg) & $\alpha_{\rm sp}$ & $188^{+31}_{-34}$ \\
Typhon rotation pole decl. (deg) & $\delta_{\rm sp}$ & $77.2^{+4.3}_{-4.0}$ \\
\enddata
\tablecomments{Upper table refers to parameters in our fits, while the lower portion shows derived parameters. Reported values for parameter posterior distributions show the median, 16th, and 84th percentile values. For Echidna's mass and Typhon's rotation rate we report 84th/16th percentile upper/lower limits, respectively as their posteriors peak very close to the parameter lower/upper bounds. All fitted angles are relative to the J2000 ecliptic plane and referenced to Typhon-centric JD 2454000.0 (2006 September 21 12:00 UT). R.A. and decl. values are relative to the J2000 equatorial plane. }
\end{deluxetable}

\begin{figure*}[!ht]
    \centering
    \begin{subfigure}[b]{0.45\textwidth}
        \includegraphics[width=\columnwidth,trim={2mm 3mm 2mm 2mm},clip]{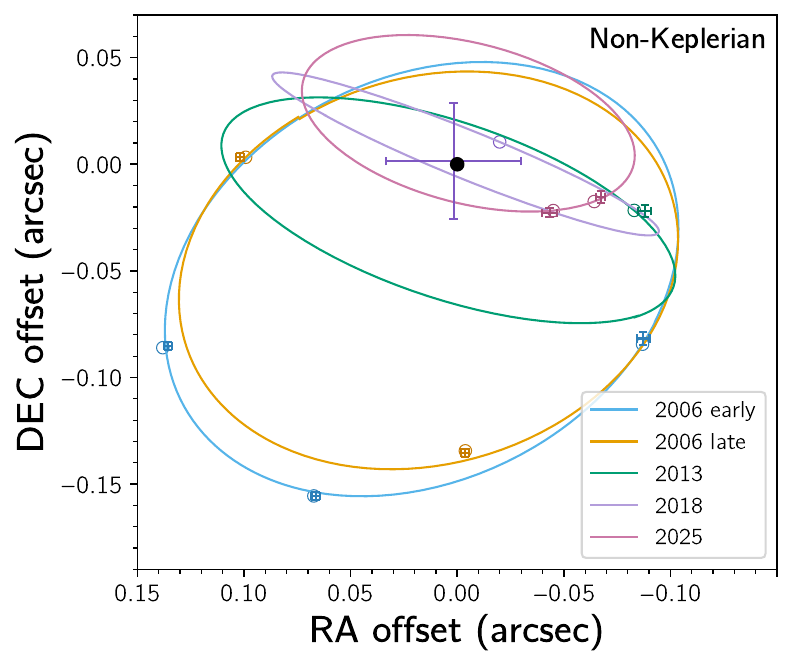}
    \end{subfigure}
    \hspace{-0.75em}
    \begin{subfigure}[b]{0.45\textwidth}
        \includegraphics[width=0.91\columnwidth,trim={2mm 3mm 2mm 2mm},clip]{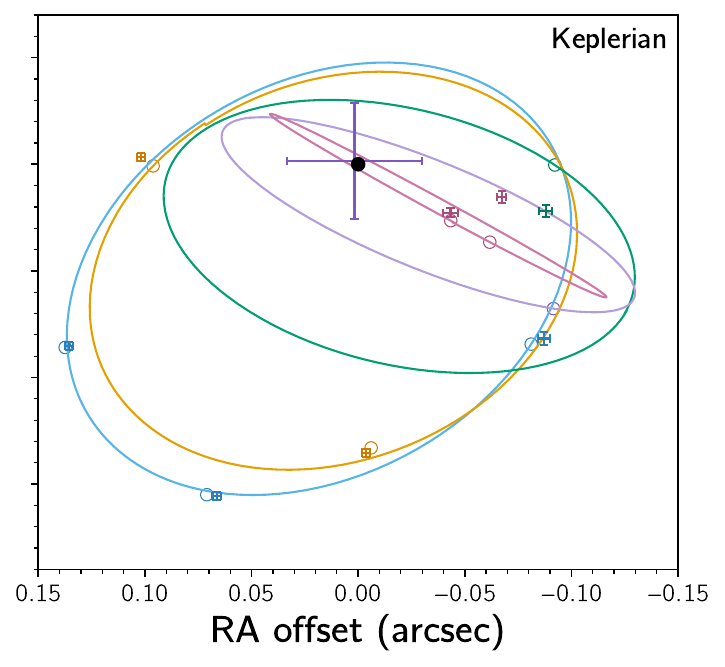}
    \end{subfigure}
    \caption{The on-sky evolution of Echidna's orbit from 2006--2025 in our non-Keplerian (left) and Keplerian (right) orbit fits. Colored lines show the on-sky projection of the orbit during different epochs. Open circles show the predicted positions of Echidna against the measured astrometry. The non-Keplerian orbit fit shows good agreement with the observations, while the Keplerian orbit fit has substantial residuals---especially for the 2013-2025 observations. }
    \label{fig:orbit}
\end{figure*}

\subsection{Orbit Results}
The output posterior distributions for each of our fitted parameters, as well as a variety of derived parameters, are shown in Table \ref{tab:orbitfit}. We also display the full 12-dimensional posterior distribution as a corner plot \citep{corner} in Figure \ref{fig:corner}. In comparison with our preliminary Keplerian fits, which achieved a best-fit $\chi^2=185$ with 11 degrees of freedom, our non-Keplerian fits achieved a best-fit $\chi^2=11$ with six degrees of freedom for a $\chi^2_\nu$ ($\chi^2$ per degree of freedom) of 1.8. Although slightly elevated, the probability of achieving such a value (or worse) due purely to random chance (e.g., the $\chi^2$ p-value) is 9\%. As such, we consider the spin-orbit model presented to provide a good representation of the system given the current data.

The precession caused by Typhon's nonspherical shape can be seen in Figure \ref{fig:orbit}. There, we show the observations compared with the model on the sky for both our best non-Keplerian and Keplerian orbit fits. While both the Keplerian and non-Keplerian fits are somewhat in agreement on the orbit through the end of 2006, non-Keplerian precession quickly alters the orbit. This is first seen in 2013, where the observation made from Keck appears to be significantly discrepant with the Keplerian orbit plane. The single epoch observation from HST in 2018, which was a non-detection, was coincidentally consistent with the 2006 orbit plane, but was at the wrong orbit phase. Finally, the two HST observations in 2025, taken close to a single orbit period apart, revealed a significantly precessed orbit plane. 

\section{Discussion}
\label{sec:discussion}

\begin{figure}
    \includegraphics[width=\columnwidth,trim={2mm 3mm 2mm 2mm},clip]{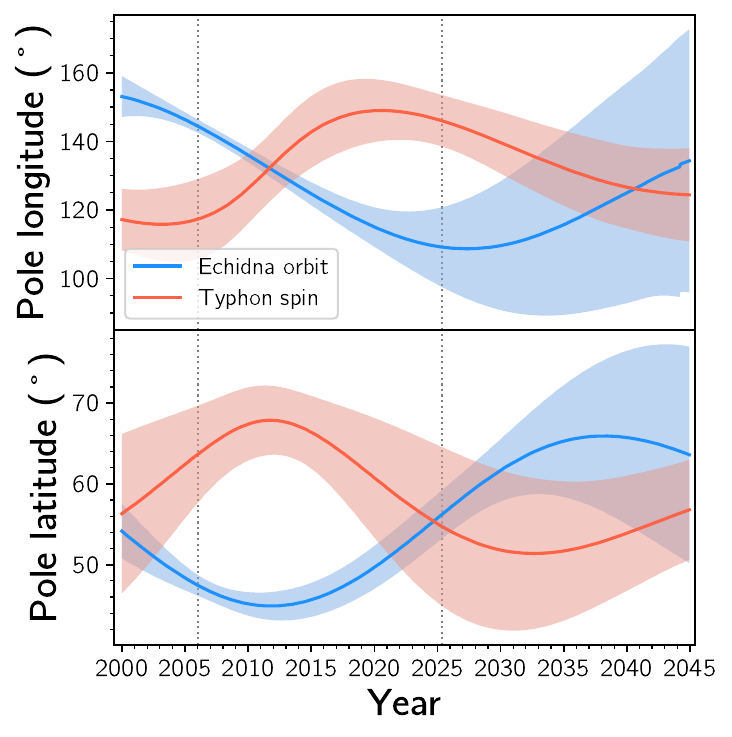}
    \caption{Precession of Typhon's spin pole and Echidna's orbit pole. Colored lines and shaded regions indicate the mean and standard deviation of the orbit/spin pole direction of a sample of 500 posterior draws. \added{Vertical dotted lines show the times of the first and last of our observations.} Coordinates are referenced to the J2000 ecliptic reference frame. Since we assume all angular momentum in the system is contained within these two sources, any single self-consistent sample of parameters would produce perfectly sinusoidal variations. The angular separation between the red and blue curves is equal to the $23.8^{+5.1}_{-4.6}\degr$ misalignment between Typhon's spin and Echidna's orbit.}
    \label{fig:latlon_evolution}
\end{figure}

\subsection{Spin-Orbit Evolution of Typhon and Echidna}
\label{sec:sporb}

Consistent with the strong rejection of the Keplerian orbit fit, our non-Keplerian fit detects Typhon's gravitational harmonic $J_2R^2=726^{+87}_{-73}$---a nearly $10\sigma$ detection. To date, this is the most confident detection of $J_2$ for any TNB, even for large TNOs with dozens of observations \citep{proudfoot2024bpm3,proudfoot2025beyond}. Our orbit fits show substantial precession with (orbit averaged) rates of $\dot{\omega}=3.5^{+0.8}_{-0.7}\degr$ yr$^{-1}$ and $\dot{\Omega}=-4.7^{+0.5}_{-0.9}\degr$ yr$^{-1}$. These yield precession periods of $P_{\omega}=102^{+22}_{-20}$ yr and $P_{\Omega}=76^{+14}_{-8}$ yr. 

Our model is also able to approximately recover the pole orientation of Typhon (at the reference epoch). Besides Haumea, whose two precessing moons reveal the pole orientation \citep{proudfoot2024bpm3}, this is the first dynamical determination of a rotation pole of a TNO. Precession of other TNBs has been detected \citep[e.g.,][]{nelsen2025beyond}, but in those cases only the spin-orbit misalignment was able to be measured, defining a double-sided cone of possible rotation poles.  

In an interacting spin-orbit system like Typhon, the nodal precession of the secondary's orbit is coupled with the axial precession of the primary. So, as Echidna's orbit slowly precesses, Typhon's rotation axis will also precess, conserving angular momentum (neglecting any small effects from the Sun or other planets). We show the coupled spin-orbit evolution of the system in Figure \ref{fig:latlon_evolution}. 

Both Typhon's rotation axis and Echidna's orbit pole precess around the combined direction of total angular momentum ($L_{\rm tot}$). The relative amplitude of the precession is set by the ratio $\eta=L_{\rm sp}/L_{\rm orb}$, where $L_{\rm sp},L_{\rm orb}$ are the spin and orbit angular momenta. For $\eta\ll 1$, angular momentum is concentrated in the mutual orbit, producing low amplitude nodal precession but high amplitude axial precession. On the other hand, for $\eta\gg 1$, angular momentum is concentrated in the spin of the primary, producing high amplitude nodal precession and low amplitude axial precession. Typhon, with $\eta\sim1$, sits between these two edge cases, exhibiting significant axial \textit{and} nodal precession. We discuss the possibility of detecting axial precession with rotational light curves in Section \ref{sec:lc}. 

The spin-orbit misalignment between Typhon's spin and Echidna's orbit is substantial, $\epsilon=23.8^{+5.1}_{-4.6}\degr$, which strongly enhances the detectability of the nodal precession. This misalignment can be thought of in two ways, the obliquity of Typhon relative to Echidna's orbit or Echidna's inclination relative to Typhon's equator. Perhaps this misalignment is unsurprising, as Echidna's eccentricity is also excited ($e=0.53\pm0.01$). Unfortunately, relatively few other TNB spin-orbit alignment determinations have been made, making comparison with other systems difficult. 

\citet{nelsen2025beyond} studied the Altjira system in detail, revealing that one of the binary components is likely a close/contact binary with $\epsilon=14^{+12}_{-6}\degr$. Borasisi-Pabu was similarly reported to have at least one contact binary component with $\epsilon=45^{+7}_{-5}\degr$ inclined to its mutual orbit \citep{proudfoot2024bpm2}, although new observations may indicate systematic issues with the previous determination (B. Proudfoot et al., in preparation). The moons of some larger TNOs are inferred to be on near equatorial orbits ($\epsilon\approx0\degr$) based on their mutually tidally locked states \citep{christy1978satellite,szakats2023tidally,bernstein2023synchronous,collyer2025synchronous}. Haumea's smaller moon Namaka and Quaoar's larger moon Weywot appear to have significant inclinations \added{(Namaka, $\sim13\degr$; \citealp{rb09,proudfoot2024bpm3}; Weywot, $\sim5\degr$; \citealp{proudfoot2025beyond})} that may be related to past dynamical excitation \citep{cuk2013dynamics}. With further $\epsilon$ measurements among other TNBs of similar size to Typhon--Echidna, greater clarity will be provided on the distribution of spin-orbit alignments among TNBs.

Although we do not explicitly include Echidna's rotation in our model, Echidna will also undergo significant axial precession, the exact details of which are subject to Echidna's size, shape, and rotation period. Using representative orbital parameters from our model, we ran a suite of integrations that include Echidna's spin properties. We used the size of Echidna determined by thermal measurements \citep{santos2012tnos} and a shape similar to that which we find for Typhon (see section \ref{sec:shape}) which is typical for small, icy bodies. In our series of integrations, we varied Echidna's pole orientation and rotation period within reasonable values to explore the variety of spin evolution.

\added{When Echidna's orientation is approximately aligned with its orbit ($\lesssim15\degr$) and has a} rotation period $\lesssim45$ hours, Echidna's rotation pole rapidly precesses around its orbit pole with precession periods of $\lesssim10$ years. Higher obliquity values increase the precession period, with $P_{\rm prec}\propto 1/\cos{\epsilon}$ \citep[compare with][]{hastings2016short}. At rotation periods $\gtrsim$45 hours, Echidna's rotation becomes chaotic, with periodic kicks near its closest approach to Typhon causing tumbling. This marked transition to a chaotic regime is due primarily to Echidna's eccentric orbit, which leads to spin-orbit resonance overlap (\citealp{wisdom1984chaotic}, see also Section 5.7 of \citealp{1999ssd..book.....M}). These spin-orbit simulations also show that Echidna's spin has no effect on Typhon's spin or the mutual orbit---even when chaotically rotating.

With resolved photometric observations of Echidna, it may be possible to confirm rapid axial precession or chaotic rotation, but the extremely small on-sky separation ($<90$ mas) makes these observations quite difficult. Detecting Echidna's photometric variability in unresolved photometric observations \textit{may} be possible with enough high-precision photometry \citep[like done for Haumea and Hi`iaka,][]{fernandez2025accurate}, but the fading brightness of the system ($V\sim22$ in 2025) as it moves outwards in its heliocentric orbit will complicate these efforts. 

\subsection{The Typhon--Echidna Mass Ratio}
\label{sec:mr}

One notable feature of our orbit fits is that they break the Keplerian mass degeneracy, in which the individual component masses cannot be independently measured. This arises as a direct consequence of the complex spin-orbit interactions at play in the system. The non-Keplerian precession rates of a binary are a function of $\epsilon$, with $\dot{\Omega}\propto J_2\cos{\epsilon}$ and $\dot{\omega}\propto J_2\sin^2{\epsilon}$ \citep[see][]{scheeres2000evaluation,proudfoot2024bpm2}. Measurement of both precession rates therefore uniquely defines $\epsilon$ and $J_2$. For nodal precession observed over a sufficiently long baseline, we can additionally determine the pole about which the orbit precesses, corresponding to the direction of the system’s total angular momentum, $L_{\rm tot}$. In our simplified spin-orbit model, in which angular momentum is only contained within the mutual orbit ($L_{\rm orb}$) and the primary's spin ($L_{\rm sp}$), the direction of $L_{\rm tot}$ lies somewhere between $L_{\rm orb}$ and $L_{\rm sp}$, weighted by their relative magnitudes. Since $L_{\rm orb}\propto(M_1M_2)/(M_1+M_2)$ and $L_{\rm sp}\propto M_1$, determining the precession rates and precession pole can break the mass degeneracy. With a two decade observational baseline---about $25\%$ of the nodal precession period---our orbit fits are able to recover the direction of the total angular momentum, and thereby break the mass degeneracy.

As can be seen in Figure \ref{fig:corner}, our fits provide only a weak detection of Echidna's mass, yielding a mass ratio, $q=M_e/M_t=0.068^{+0.078}_{-0.046}$. This is somewhat lower than expected; under the assumption of spherical bodies with equal densities and albedos, a size ratio of $0.55\pm0.015$ \citep{grundy200842355} implies $q=0.17\pm0.01$, just outside our upper uncertainty bound. In reality, Typhon and Echidna may not have equal densities, nor equal albedos, and clearly do not have spherical shapes. Direct size and shape measurements from stellar occultations would constrain possible albedo differences and, in turn, allow the relative densities of the two bodies to be constrained.

We emphasize that this constraint should be interpreted as a dynamical limit rather than a precise measurement of the mass ratio; nevertheless, it represents the first TNB mass ratio inference based solely on non-Keplerian spin–orbit dynamics in a two-body system. To date, only a few other TNBs have mass ratio constraints derived independently of an assumed density. These measurements have generally been made through detection of the barycentric motion of the primary \citep{brozovic2015orbits,brown2023masses} or through dynamical constraints imposed by a third component \citep{rb09,ragozzine2024beyond,proudfoot2024bpm3}. Exploring the mass ratios of small binaries like Typhon--Echidna will allow us to better understand the formation and early dynamical evolution of TNBs.

As expected, we find that the mass of Echidna is correlated with both Typhon's spin rate and pole orientation, as these parameters can be varied while still maintaining the same net direction of the total angular momentum. In our fits, we limited exploration of Typhon's rotation period to be between 5 and 20 hours. At rotation periods greater than 20 hours, corresponding to a decrease in $L_{\rm sp}$, the inferred mass of Echidna must shrink to compensate, reaching unreasonably small values ($q<0.01$). We therefore infer that Typhon’s true rotation period is likely $<20$ hours, unless Typhon and Echidna have substantially different albedos and/or densities. Detailed study of Typhon's rotational light curve should further constrain the system's mass ratio.

In addition to the mass ratio, we derive a total system mass of $M_{\rm sys}=1.13^{+0.05}_{-0.05}\times10^{18}$ kg, exceeding the value reported by \citet{grundy200842355} at the $3\sigma$ level. This difference is likely due to our inclusion of precession. With a system mass, we can derive a system density slightly higher than past estimates, $\rho_{\rm sys}=440\pm100$ kg m$^{-3}$ when using the effective thermal diameter \citep{santos2012tnos}. This points to an extremely porous body, like other TNOs of similar size. Further improvement in the density of the system should focus on observing stellar occultations by both Typhon and Echidna.

\added{We note that the changing aspect angle (see Section \ref{sec:lc}) may significantly alter the thermal emission and thermal light curve of Typhon. \textit{Spitzer} and \textit{Herschel} data of Typhon were taken in 2008 and 2010, respectively, when the aspect angle was between $\sim$72-77$\degr$. With a nearly equator-on orientation, thermal emission was likely minimized, potentially making the \citet{santos2012tnos} thermally-derived size an underestimate of Typhon's true size. Further, evolution between the \textit{Spitzer} and \textit{Herschel} observations, although only $\sim5\degr$, could bias the resulting measurements. Future work should explore thermophysical models that directly incorporate the shape and orientation of Typhon along with self-consistent evolution of the thermal emission. This may provide another method to better measure Typhon's density.}


\subsection{Typhon's Shape}
\label{sec:shape}

With a confident, high-precision measurement of Typhon's $J_2R^2$, we are able to make inferences about its shape given an assumed shape model. The choice of a shape model is a critical assumption when inverting a $J_2R^2$ measurement as multiple shapes can provide identical values. Including additional information, like light curves or occultations, can certainly improve models, but both can have significant drawbacks \citep[e.g.,][]{harris2020asteroid}. So, for this section, we focus on the simplest shape model available, homogeneous triaxial ellipsoids, \added{although we do note that departures from either an ellipsoidal shape and homogeneous density structure are expected for a body of Typhon's size.}

Given a non-zero measurement of $J_2R^2$, we can immediately infer that the ellipsoid semi-axes ($a,b,c$) cannot be equal. Further, light curve observations of the system do provide a $\Delta m = 0.07\pm0.01$ mag \citep{thirouin2010short}, implying $a\ne b$ if Typhon's surface has a uniform albedo. Although the light curve observations do not provide a period with high significance, we assume that the derived $\Delta m$ approximates the true light curve amplitude of the system.

Under the assumption of a homogeneous triaxial ellipsoid ($a>b>c$), we have three equations that govern the shape of Typhon. First, the relationship between $J_2R^2$ and $a,b,c$ is
\begin{align}
    J_2R^2 = \frac{1}{10} \left( a^2 + b^2 - 2 c^2 \right)
\end{align}
\noindent \citep{1995geph.conf....1Y}. Second, the light curve amplitude is\begin{align}\label{eqn:rlc}
  \Delta m = -\frac{5}{2} \log\left[\left(\frac{b}{a}\right)\left(\frac{(a/c)^2\cos^2\theta+\sin^2\theta}{(b/c)^2\cos^2\theta + \sin^2\theta}\right)^{1/2}\right]
\end{align}
\noindent where $\theta$ is the polar aspect angle defined by the angle between the Earth-Typhon vector and Typhon's rotation pole. When the observations of Typhon's light curve were acquired in 2003, $\theta=61^{+6}_{-8}\degr$. To account for systematic uncertainties in the light curve amplitude, we double the uncertainty in $\Delta m$ to get a more realistic uncertainty in the shape model. 

Lastly, we require the ellipsoid's surface area to match the derived thermal diameter of Typhon, $D_{\rm surf}=162\pm7$ km. Again, we double the uncertainty in this measurement to account for systematic uncertainties. The surface area of an ellipsoid has no closed form solution, but is well approximated by
\begin{align}
    S \approx 4\pi \left( \frac{a^pb^p + a^pc^p +b^pc^p}{3} \right)^{1/p}
\end{align}
\noindent where $p=1.6075$ \citep{Thomsen_2004}.

Numerically solving those three equations for a statistical sample of orbit solutions \added{(i.e., samples from the $J_2R^2$ posterior)}, we find $a=93^{+8}_{-6}$ km, $b=84^{+6}_{-6}$ km, and $c=65^{+9}_{-8}$ km. In terms of axis ratios, we find $c/a=0.70^{+0.05}_{-0.06}$ and $b/a=0.90^{+0.03}_{-0.04}$. 

This shape model, although significantly flattened and elongated, is fairly typical for its size range and dynamical population (see, \citealp{tegler2005period,rousselot2021new,pereira2024physical,rizos2024study}, for a review, see \citealp{fernandez2025Centaur}). The shape is also very similar to Chariklo's \citep[$c/a=0.69^{+0.04}_{-0.02}$, $b/a=0.94^{+0.01}_{-0.02}$,][]{leiva2017size,morgado2021refined}, which has been determined with the use of stellar occultations. 

\added{We also draw comparison with Arrokoth's larger lobe, which has a $c/a=0.68$ \citep{2026arXiv260530069P}, as Typhon likely formed in a similar streaming instability collapse process which is thought to have formed Arrokoth's individual lobes before their eventual merger \citep{mckinnon2020solar}.}

\added{Since most small TNOs have poorly measured sizes, determining the \jt{} of a body alone is difficult since the uncertainty will be strongly dominated by the size uncertainty. However, with a well-determined size and shape model,} we can also determine the $J_2$ coefficient alone, $J_2=0.114^{+0.025}_{-0.021}$. Again, few \jt{} measurements have been made for TNBs of a similar size, and those that have are large \citep[e.g.,][]{nelsen2025beyond}, likely due to observational biases related to the ease of detecting rapid precession \citep{proudfoot2024bpm2}. Comparing with asteroid binaries, we find that our \jt{} measurement is fairly typical of similar sized objects \citep{marchis2005mass,2011AJ....141..154F,marchis2014puzzling,minker2025dynamical,minker2025origin}.

\begin{figure*}
    \centering
    \begin{subfigure}[b]{0.49\textwidth}
        \includegraphics[width=\columnwidth,trim={2mm 3mm 2mm 2mm},clip]{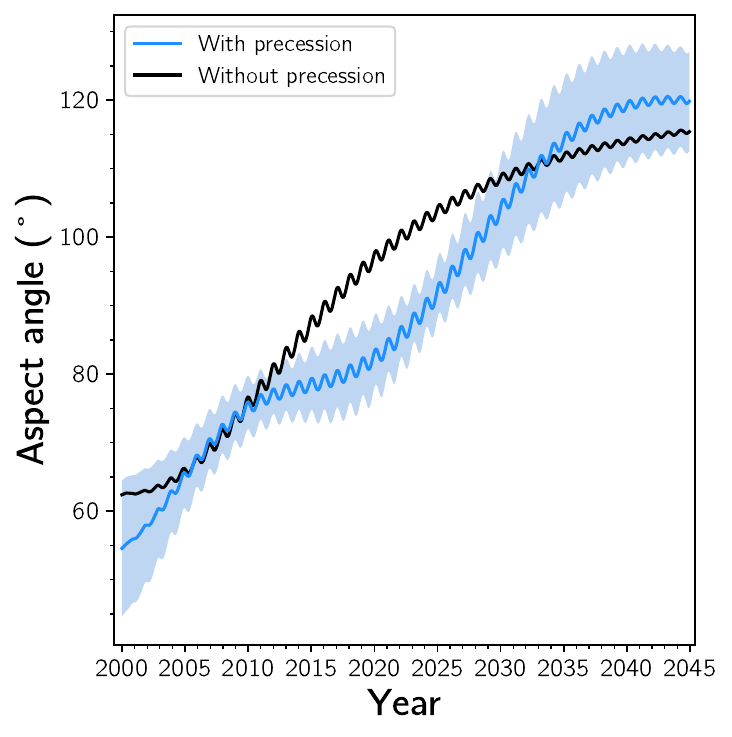}
    \end{subfigure}
    \hspace{-0.25em}
    \begin{subfigure}[b]{0.49\textwidth}
        \includegraphics[width=\columnwidth,trim={2mm 3mm 2mm 2mm},clip]{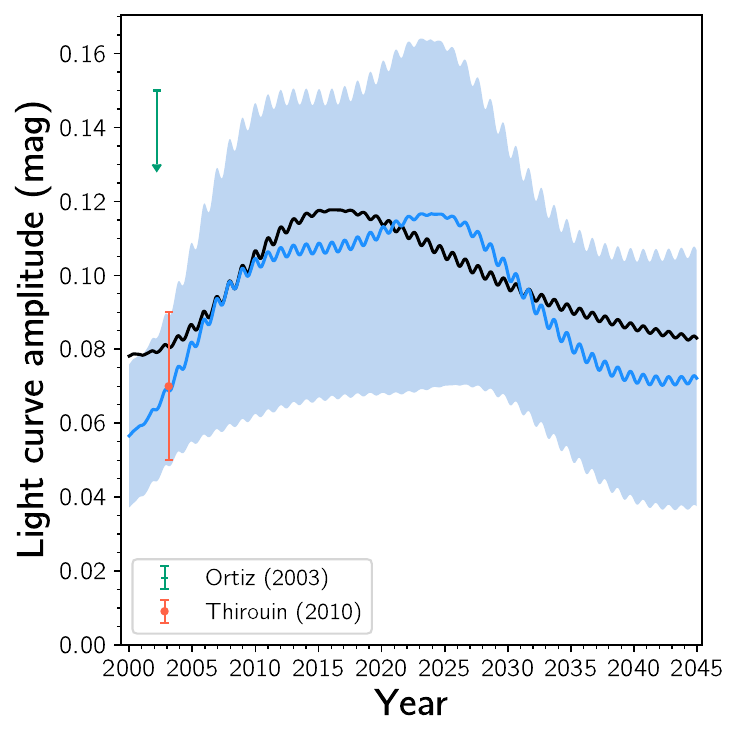}
    \end{subfigure}
    \caption{Evolution of Typhon's aspect angle (left) and light curve amplitude (right). Aspect angle is defined as the angle between the Earth-Typhon vector and Typhon's rotation pole. Light curve amplitude assumes a triaxial shape model as detailed in Section \ref{sec:shape}. Blue lines and shaded regions show the mean and standard deviation of 500 posterior draws. \added{Vertical dotted lines show the times of the first and last of our astrometric observations.} Black lines show the case with no precession---referenced to the orbit fitting epoch and using the median shape model---illustrating the evolution of the aspect angle due to Typhon's heliocentric orbit and Earth's orbit (the annual variation). Light curve amplitude measurements/upper limits are from \citet{ortiz2003study} and \citet{thirouin2010short}, the latter of which was used to construct the shape model explaining the good fit at that epoch.}
    \label{fig:spin_evolution}
\end{figure*}

We emphasize this shape model is built on \textit{many} assumptions; for example, we assume uniform surface albedo of Typhon, equal albedos between Typhon and Echidna, homogeneous interior structure of Typhon, and no precession contribution from Echidna, among others. If any of these assumptions are untrue, the derived shape model can change substantially. Further validation (or rejection) of this shape model can be accomplished with a variety of techniques, most notably, stellar occultations and light curve monitoring.

Stellar occultations are the gold-standard for probing the sizes and shapes of TNOs without expensive spacecraft missions. Typhon currently has relatively fast on-sky motion, thanks to its location near perihelion, offering numerous opportunities to observe occultations. Although its heliocentric ephemeris is somewhat uncertain ($3\sigma$ uncertainty is $\sim$100 mas), a dedicated series of occultation campaigns could feasibly study Typhon's size and shape. In addition, due to the small angular distance to Echidna ($<0.09$ arcsec, see Figure \ref{fig:orbit}), Echidna's shadow should often occult the same star, although the shadow paths will not necessarily be coincident. Observations of both Typhon and Echidna during occultations can also be used to further improve the system's mutual orbit solution \citep[e.g.,][]{rommel2025stellar,proudfoot2025beyond}. In a forthcoming publication, we will use the orbit fits we present here---along with other orbit fits from the \textit{Beyond Point Masses} project---to provide a robust methodology for predicting stellar occultations by TNBs (F. Rommel et al., submitted). 

\subsection{Light Curve Evolution}
\label{sec:lc}

Another technique to validate our shape model is long-term monitoring of Typhon's light curve. As Typhon moves around its heliocentric orbit, its aspect angle evolves significantly over time. Typhon's axial precession adds an additional source of modification of the aspect angle. Using a sample of our posteriors, in the left panel of Figure \ref{fig:spin_evolution}, we show the evolution of Typhon's aspect angle over 40 years, compared with the aspect angle assuming no axial precession. Although the modification is relatively modest given $\epsilon=23.7^{+5.1}_{-4.6}\degr$, the change in the aspect angle is significant and capable of producing rapid changes over short timescales.

Using Equation \ref{eqn:rlc} and the shape model we derived earlier, we can also determine how Typhon's light curve amplitude evolution is modified by precession (right panel of Figure \ref{fig:spin_evolution}). The no precession case is within the $1\sigma$ uncertainty provided by our model including precession (and using the shape model discussed previously), but has a substantially more complicated evolution. The uncertainty in the light curve amplitude is mostly due to uncertainties in Typhon's rotation rate---which significantly alters the precession amplitude---although both the size and light curve amplitude also contribute to the substantial uncertainties.

With enough high precision light curve measurements ($\sigma_{\Delta m} \lesssim0.01$ mag), our spin-orbit model could be substantially improved---due to both the more rigorous period determination and the constraints provided by the light curve amplitude. Unfortunately, to our knowledge, no photometry of Typhon taken since 2003 has been published. Future observations, or light curves from archival sources, will thus be necessary to further constrain our spin-orbit-shape model. Ideally, with enough light curve observations, both the orbit and the light curve evolution could be fit jointly. More generally, spin-orbit-shape modeling can be applied to many TNO binaries, not just Typhon--Echidna. We explore these in M. Thatcher et al., in preparation.

\added{Although targeted light curve measurements are ideal, sparse photometry from all-sky survey---like that to be conducted with the Vera Rubin Observatory's Legacy Survey of Space and Time (VRO/LSST)---can also play a role in measuring Typhon's light curve evolution. Over the 10-year LSST program, significant light curve evolution may be detected (see Figure \ref{fig:spin_evolution}), providing a way to test our models. Although sparse photometry of a rapidly varying light curve may be challenging, with enough measurements, currently available tools are able to detect such light curve changes \citep[e.g.,][]{2025PSJ.....6..280R}.}

\begin{figure}
    \includegraphics[width=\columnwidth,trim={3mm 3mm 2mm 2mm},clip]{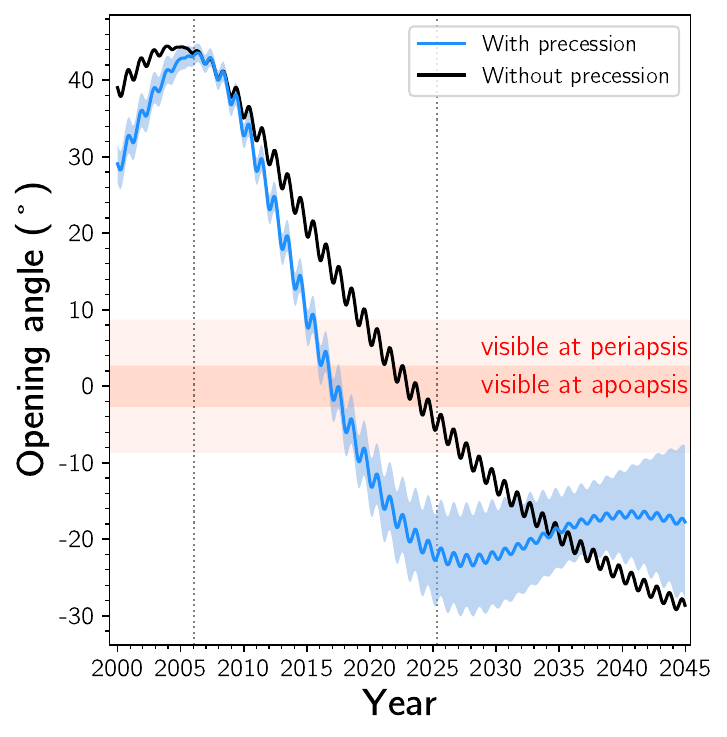}
    \caption{Evolution of the opening angle of the orbit with and without precession. Opening angle is defined as the angle between the Earth-Typhon vector and the mutual orbit normal. Mutual events occurring at/near periapsis are visible with opening angles between maximum of $\pm8.7\degr$ (light red region), while events at/near apoapsis are visible between $\pm2.7\degr$ (darker red region). }
    \label{fig:op_evolution}
\end{figure}

\subsection{Mutual Events}
\label{sec:me}
Based on the orbit solution derived from only the 2006 observations, Typhon--Echidna were predicted to enter a mutual event season in $\sim$2019 lasting until $\sim$2026 \citep{grundy200842355}. However, with the substantial precession which our non-Keplerian orbit fits have revealed, the mutual event season has substantially shifted. In Figure \ref{fig:op_evolution}, we show the evolution of the orbit opening angle---defined as the angle between the Earth-Typhon vector and the mutual orbit normal. At low opening angles, mutual events are visible as Typhon and Echidna occult, eclipse, and/or shadow each other. Unfortunately, the mutual orbit's precession accelerated the evolution of the opening angle significantly, with mutual events visible from $\sim$2015 until $\sim$2019. 

Although the missed opportunity to catch mutual events is indeed disappointing, precession does provide some benefit. The orbit plane of a prograde mutual orbit (where the heliocentric and mutual orbit normals are in the same hemisphere) will regress ($\dot{\Omega}<0$), increasing the apparent rate at which the mutual orbit plane appears to move. This allows mutual events to occur more frequently than would be assumed from a simple Keplerian orbit. The period between mutual event seasons ($P_{\rm me}$) is given by
\begin{align}
    \frac{1}{P_{\rm me}} = \frac{1}{2} \left| \frac{360\degr}{\dot{\Omega}} - \frac{360\degr}{n_{\sun}} \right|
\end{align}
\noindent where $n_{\sun}$ is Typhon's heliocentric mean motion. For Typhon, given our measured $\dot{\Omega}\sim-5\degr$ yr$^{-1}$ and an orbit averaged $n\sim360\degr/P_{\rm orb}$, we find $P_{\rm me}\sim30$ yr. If the orbit was instead retrograde (resulting in $\dot{\Omega}>0$), $P_{\rm me}\sim50$ yr. Since Typhon is on a high-eccentricity heliocentric orbit ($e_{\sun}=0.536$), $n_{\sun}$ varies considerably over its orbital period, making this more complex. However, as can be seen in Figure \ref{fig:op_evolution}, it is possible that the system will again enter a phase of mutual events in the mid-to-late 2040s. 

In other TNBs, precession will also play a role in the timing of mutual events, with strong dependence on the orbit orientation and precession rate. For prograde binaries, any precession rate will cause mutual events to occur more frequently than twice per heliocentric orbit period. On the other hand, outcomes can vary for retrograde binaries. For binaries with $\dot{\Omega}<2n_{\sun}$, the time between mutual event seasons increases (above that for the no precession case). When the two rates are commensurate, $\dot{\Omega}=n_{\sun}$---called the evection resonance---mutual events will never be visible from Earth (or will always be visible). For $\dot{\Omega}>2n_{\sun}$, the time between seasons again is less than half the orbit period. 

To understand these effects on TNBs in general, we first need to estimate reasonable precession rates of TNBs based on the population of Keplerian orbit fits. Typical TNBs have their orbit angles measured---in high quality Keplerian orbit fits---with $\lesssim5\degr$ precision over an observational baseline of $\sim$5 yrs \citep{proudfoot2024bpm2}. By assuming the accumulated precession over that baseline is less than the uncertainty on the orbit angles, we can crudely estimate a typical $\left|\dot{\Omega}\right|\lesssim1\degr$ yr$^{-1}$. Typical TNBs with heliocentric semi-major axes of $\sim45$ au will have $n_{\sun}\sim1-1.5\degr$ yr$^{-1}$. 

For TNBs on prograde mutual orbits, any precession will increase the frequency of mutual events. However, for retrograde TNBs with typical $\dot{\Omega}<2n_{\sun}$, mutual events will be less often than twice per heliocentric orbit. Fortunately, TNBs predominantly have prograde orbits at a ratio of $\sim$4:1 \citep{grundy2019mutual}, so this will not significantly affect the occurrence of TNB mutual events. 

Precession can also change the duration of a mutual event season. As seen in Figure \ref{fig:op_evolution}, the duration of the season was significantly reduced by the precession. Alternatively, for retrograde orbits, the season can be extended. In addition to the nodal precession effects, apsidal precession of eccentric mutual orbits can alter the duration of the season by changing the range of opening angles that events can be seen at. The combination of these effects can be quite complex, emphasizing the need for full non-Keplerian orbit solutions when predicting mutual events \citep[see][]{proudfoot2026trans}.

\subsection{Tidal Evolution}
\label{sec:tide}

As a relatively tight TNB, Typhon--Echidna are expected to undergo tidal evolution, which can affect both Typhon and Echidna's spins, as well as the mutual orbit $a$ and $e$. Tides are notoriously difficult to model and dependent on a huge range of quantities that are mostly unconstrained. However, simple order of magnitude estimates can be useful in understanding binaries. First, looking to tidal despinning, the timescale for despinning Typhon is
\begin{align}
    \tau_{\rm sp,1} = \Delta\omega_{1}\frac{Q_1'M_1a^6}{GM^2_2R_1^3}
\end{align}
\noindent where $\Delta\omega$ is the change in angular rotation from an initial rotation rate to the synchronous state, and $M_1,M_2,R_1,R_2$ are Typhon and Echidna's masses and radii, respectively. Switching subscripts gives the equation for the despinning timescale of Echidna. The modified tidal quality factor $Q'$ is the tidal quality factor $Q$ corrected for the effects of rigidity, and is
\begin{align}
    Q_i'=Q\left(1+ \frac{19\mu}{2g_i\rho_i R_i} \right)
\end{align}
\noindent where $\mu$ is the rigidity, $g$ is the surface gravity ($g=GM/R^2$), and $\rho$ is the density. The rigidity of Typhon and Echidna are unknown, so we take a range of allowable values from $10^6$ Pa (like that of unconsolidated sand) up to $4\times10^{9}$ Pa (that of solid ice). Likewise, $Q$ is unknown, but is typically taken as $\sim100$ for small, icy bodies. 

To place a lower limit on the despinning time of Typhon and Echidna, we use the lower bound on rigidity and $Q=10$, giving $\tau_1\sim60$ Myr and $\tau_2\sim5$ Myr. Then for $\mu=4\times10^9$ Pa and $Q = 100$, $\tau_1>10^3$ Gyr and $\tau_2\sim200$ Gyr. Since Typhon appears to still be rotating much faster than the binary's mutual orbit \citep{thirouin2010short}, we can infer that tides are indeed weaker than our lower bounds. Unfortunately, however, with $Q=100$, we need only $\mu\sim10^7$ Pa for $\tau_1>4.5$ Gyr, only an order of magnitude more rigid than unconsolidated sand, so little can be determined about the true tidal parameters of the system. 

Looking to the mutual orbit eccentricity, following \citep{1999ssd..book.....M}, the timescale for circularization of the orbit is
\begin{align}
    \tau_{\rm circ} = \frac{4}{63} \frac{M_2}{M_1} \frac{a^5}{R_2^5} \frac{Q_2'}{n}
\end{align}
\noindent where $n$ is Echidna's mean motion. Assuming rapid tidal evolution using $\mu=10^6$ Pa and $Q=10$, $\tau_{\rm circ}\sim10$ Myr, while for $\mu=4\times10^9$ Pa and $Q = 100$, $\tau_{\rm circ}\sim300$ Gyr. Again, given the eccentric mutual orbit, we can rule out rapid tidal evolution on Myr timescales, but cannot place meaningful constraints on $Q$ or $\mu$. 

\added{The excited orbital state, even given the compact nature of the mutual orbit, appears to be consistent with modeling of coupled Kozai cycles with tidal friction \citep{porter2012kctf}. When including the effects of \jt, such models predict that compact binaries can retain substantial eccentricity over the age of the solar system. As such, it is entriely possible that given Typhon's shape and tidal parameters, it may have undergone relatively little tidal evolution.}

\subsection{Origin of the Typhon--Echidna System}
\label{sec:origin}

Although the role of tides cannot be fully determined, it is interesting to consider how the evolution of the system's heliocentric orbit could have contributed to recent evolution of the binary. Typhon--Echidna is on a giant planet crossing orbit, with its perihelion interior to Uranus' orbit, allowing frequent encounters with the giant planets. \citet{araujo2018journey} explored the evolution of Typhon and found that encounters with giant planets are common. When encountering giant planets, binary systems like Typhon can have their mutual orbits substantially excited, with the majority of binaries being disrupted over their lifetimes as Centaurs \citep{brunini2014dynamical}. Previous works have suggested that Typhon--Echidna have a $\sim10\%$ chance that encounters with giant planets should have disrupted the binary by now, with a considerably higher chance that the binary could have been at least excited \citep{noll2006discovery,stansberry2008physical}. Given the possibility of past orbital excitation, the excited state of Typhon and Echidna may not be due to relatively ineffective tides, but instead could be the result of a recent ($\sim$Mya) close encounter with a giant planet. 

This hypothesis is testable. If Typhon--Echidna originated as a tight binary reminiscent of other TNBs (like 2003 QA$_{91}$, 2000 WK$_{183}$, 2000 OJ$_{67}$, and 2000 CM$_{114}$, all with similar mass and semi-major axis, see \citealp{grundy2019mutual}), we could expect an extensive history of tidal evolution, like that evidenced by the low eccentricities in those similar binaries. Although Typhon could not have been fully tidally evolved, given its rapid rotation rate, Echidna may have been able to be despun significantly, especially since the tidal despinning timescales of Echidna are an order of magnitude faster than Typhon. After evolution onto its current planet-crossing orbit, close encounters may have excited the binary into a high eccentricity and inclination (relative to Typhon's equator) orbit. During these encounters, the spins of Typhon and Echidna are relatively unaffected, but the increase in eccentricity can provide place Echidna into a chaotic rotation regime. 

On the other hand, if the system's mutual orbit is reflective of its formation, tides must be ineffective given the substantial mutual orbit eccentricity. Since the tidal circularization is controlled by the same geophysical properties as tidal despinning, in this scenario, Echidna cannot have undergone significant despinning and should retain its primordial spin rate, typically assumed to be $\sim$10 hours for TNOs  \citep{thirouin2014rotational}, and should not be chaotically rotating. Thus, if photometric observations of the system show Echidna chaotically rotating, it could imply the system's dynamically excited state is a relatively recent development. 

Of course, this hypothesis test carries some caveats that must be addressed. First, the role of tides in eccentricity damping can be complex, with certain situations enabling eccentricity pumping \citep{arakawa2021tidal}. The interdependence of eccentricity damping/pumping and despinning is complex \citep[e.g.,][]{goldberg2024chaotic}; further detailed spin-orbit-tidal modeling may reveal some evolutionary pathways that permit long-lived eccentric orbits and effective despinning tides. These pathways, however, must also explain the substantial spin-orbit misalignment that we observe, which may be difficult to explain with tidal interactions alone. 

Second, under this scenario, we assume Typhon originated as a close binary with a substantial history of tidal evolution. Instead, the system could have originated as a more widely separated binary with ineffective tides that was driven onto a tighter and more excited orbit, enabling Echidna to retain its primordial rapid rotation. This only complicates the case where Echidna has rapid rotation, still enabling us to infer a history of tidal evolution if Echidna is chaotically rotating. 

If borne out by photometric studies of Echidna, an inference that Typhon's current dynamical state is young would provide a rare glimpse at the evolution of binaries as they enter the Centaur population. This could provide further detail into the TNO-Centaur binary dichotomy and also explain the lack of Jupiter-family comet binaries \citep{sickafoose2025near}.



\section{Conclusions}
\label{sec:conclusions}
In this paper, we have investigated the spin-orbit dynamics of the Centaur binary Typhon--Echidna. Our main conclusions are as follows:
\begin{enumerate}
    \item Based on two decades of astrometric data, we detect non-Keplerian effects at $\sim10\sigma$ confidence, strongly rejecting a Keplerian orbit model.
    \item Typhon--Echida's spin-orbit dynamics appear to be significantly excited, exhibiting substantial orbital precession of the mutual orbit, as well as high-amplitude (and detectable) axial precession of Typhon.
    \item Our retrieved non-Keplerian parameters, combined with literature measurements of Typhon's size and light curve, imply ellipsoidal semi-axes $a=93^{+8}_{-6}$ km, $b=84^{+6}_{-6}$ km, and $c=65^{+9}_{-8}$ km, when adopting a triaxial shape model. This is similar to other Centaurs and TNOs of similar size. Occultations and continued light curve measurements of both Typhon and Echidna can refine this shape model further.
    \item Since we detect a substantial fraction of the system's nodal precession period, our fits are able to constrain the Typhon--Echidna mass ratio, implying a smaller mass ratio than would be naively expected from an equal-size, equal-albedo assumption.
    \item The substantial precession of the system has significant consequences for the evolution of both Typhon's light curve as well as the timing of the binary's mutual event season. 
    \item We suggest that the excited state of the binary may be the result of a recent ($\sim$few Mya) encounter with a giant planet. This hypothesis can be tested with resolved photometric observations of Echidna.
\end{enumerate}
Although Typhon--Echidna is just one system, the spin-orbit dynamics we explore here are applicable to many TNBs. Exploration of such non-Keplerian motion in other systems will uncover the physical characteristics of TNBs, providing insight into the formation, evolution, and fate of small, icy bodies in the outer solar system. We encourage continued observations of TNBs---especially astrometric, photometric, and stellar occultation observations---to continue to help understand these fascinating bodies.

\begin{acknowledgments}
We acknowledge two anonymous reviewers who generously provided a thoughtful review of our manuscript. We thank the BYU Office of Research Computing for their dedication to providing computing resources, without which this work would not have been possible. 

The authors wish to recognize and acknowledge the very significant cultural role and reverence that the summit of Maunakea has always had within the Native Hawaiian community. We are most fortunate to have the opportunity to conduct observations from this mountain. 

B.P. is supported by the University of Central Florida Preeminent Postdoctoral Program (P$^3$). 

This research is based on observations made with the NASA/ESA Hubble Space Telescope obtained from the Space Telescope Science Institute, which is operated by the Association of Universities for Research in Astronomy, Inc., under NASA contract NAS 5–26555. These observations are associated with program 10508, 10514, and 17707. Support for Program number 17707 was provided through a grant from the STScI under NASA contract NAS5-26555.

Some of the data presented herein were obtained at Keck Observatory, which is a private 501(c)3 non-profit organization operated as a scientific partnership among the California Institute of Technology, the University of California, and the National Aeronautics and Space Administration. The Observatory was made possible by the generous financial support of the W. M. Keck Foundation.

\end{acknowledgments}

\begin{contribution}

B.P. led the overall analysis, writing of the paper, designing/scheduling of observations, and was the principal investigator of the HST program. W.G. analyzed all observations, contributed to interpretation, and provided editing support. D.R. provided access to computing resources, contributed to interpretation, and provided editing support.

\end{contribution}

\bibliographystyle{aasjournalv7}
\bibliography{all}

\end{document}